\documentclass[aps,prd,nofootinbib,amsmath,amssymb,amsfonts,twocolumn,10pt,floatfix]{revtex4-1}

\usepackage[utf8]{inputenc}
\usepackage[T1]{fontenc}
\usepackage[english]{babel}
\usepackage{lmodern}
\usepackage{xcolor}
\usepackage{graphicx}
\usepackage{ifpdf}

\ifpdf
\else
  \PassOptionsToPackage{dvipdfmx}{hyperref}
\fi

\usepackage{hyperref}
\usepackage{capt-of}

\usepackage{bbm}
\usepackage{nicefrac}

\newcommand{\commutator}[2]{\left[ #1 ,\, #2 \right]}
\newcommand{\Dt}{D_{\!t}\,}
\newcommand{\vDx}{\vec{D}_{\!\vec{x}}\,}
\newcommand{\nablap}{\vec{\nabla}_{\!\vec{p}}\,}
\newcommand{\nablax}{\vec{\nabla}_{\!\vec{x}}\,}
\newcommand{\anticommutator}[2]{\left\lbrace #1 ,\, #2 \right\rbrace}
\newcommand{\Eqref}[1]{Eq.~\eqref{#1}}

\newcommand{\bbs}{\mathbbm{s}}
\newcommand{\bbp}{\mathbbm{p}}
\newcommand{\bbv}{\mathbbm{v}}
\newcommand{\bba}{\mathbbm{a}}
\newcommand{\bbt}{\mathbbm{t}}
\newcommand{\bbw}{\mathbbm{w}}
\newcommand{\bbe}{\mathbbm{e}}
\newcommand{\vac}{\mathrm{vac.}}

\newenvironment{rarray}[1]
{%
  \setlength{\arraycolsep}{#1}
  \begin{array}
}{%
  \end{array}
}

\let\oldvec\vec
\renewcommand{\vec}[1]{\hspace{-0.5pt}\oldvec{\hspace{0.5pt}#1}}

\graphicspath{{pictures//}{plots//}}
\allowdisplaybreaks
 
\begin{document}
%-------------------------------------------------------------------------------

\title{Pair Production in Rotating Electric Fields}
%Authors in alphabetical order
\author{Alexander Blinne}
\email{alexander.blinne@uni-jena.de}
\author{Holger Gies}
\email{holger.gies@uni-jena.de}
\affiliation{Theoretisch-Physikalisches Institut,  Abbe Center of Photonics,
Friedrich-Schiller-Universität Jena, Max-Wien-Platz 1, D-07743 Jena, Germany}
\affiliation{Helmholtz-Institut Jena, Fr\"obelstieg 3, D-07743 Jena, Germany}

\hypersetup{pdftitle={Pair Production in Rotating Electric Fields},
            pdfauthor={Alexander Blinne, Holger Gies}}

\begin{abstract}
We explore Schwinger pair production in rotating time-dependent
electric fields using the real-time DHW formalism. We determine the
time evolution of the Wigner function as well as asymptotic particle
distributions neglecting back-reactions on the electric field. Whereas
qualitative features can be understood in terms of effective Keldysh
parameters, the field rotation leaves characteristic imprints in the
momentum distribution that can be interpreted in terms of interference
and multiphoton effects. These phenomena may seed characteristic
features of QED cascades created in the antinodes of a high-intensity
standing wave laser field.

\end{abstract}

\maketitle

%-------------------------------------------------------------------------------
\section{Introduction}\label{sec:intro}
%------------------------------------------------------------------------------- 

Schwinger pair production -- or the spontaneous decay of the QED
vacuum in electric fields -- is one of the rare quantum field theory
phenomena which is inherently nonperturbative but analytically well
understood. Unfortunately, any attempt at verifying this understanding
in macroscopic electric fields is hampered by the exponentially small
production rate $\sim \exp[-\pi m^2/(eE)]$, where $m$ is the electron
mass being huge for typical laboratory field strengths $E$
\cite{Sauter:1931zz,Heisenberg:1935qt,Schwinger:1951nm}.  The rapid
development of optical or X-ray high-intensity lasers has lead to many
suggestions for schemes for a first discovery
\cite{Ringwald:2001ib,Alkofer:2001ik,Schutzhold:2008pz,Ruf:2009zz,DiPiazza:2009py,Dunne:2009gi,Monin:2009aj,Baier:2009it,Heinzl:2010vg,Bulanov:2010ei,Gonoskov:2013ada},
also including the combination of lasers and strong Coulomb fields
\cite{Muller:2008zzd,Muller:2009zzf,DiPiazza:2009yi,DiPiazza:2010kg}. 

However, while the highest field intensities
in such systems may indeed gradually approach the critical intensity,
$I_{\text{cr}} = E_{\text{cr}}^2 = \left( \frac{m^2}{e} \right)^2
\simeq 4.3\times 10^{29}\, \text{W/cm}^2$, further possible physical
processes may set in that could partly or entirely swamp a pair
production signal. In particular QED cascades of successive radiation
of accelerated charges and particle production from hard photons are
expected to occur \cite{Bell:2008zzb,Kirk:2009vk,Fedotov:2010ja,Bulanov:2010gb,Elkina:2010up,Nerush:2010fe,Nerush:2011xr,King:2013zw,Bulanov:2013cga,Bashmakov:2013iwa}, which may even
fundamentally inhibit the generation of near critical intensities. 

In view of a possible discovery of Schwinger pair production in strong laser
fields, this gives rise to a crucial question: can a QED cascade seeded by an
electron (sourced by impurities of an imperfect vacuum) be distinguished from
a QED cascade seeded by Schwinger pair production? Whereas the ensemble of
electrons arising from impurities are likely to have an isotropic initial
momentum distribution, the ensemble of Schwinger created pairs can be expected
to carry information about the directionality of the electric field that lead
to pair creation.

To quantify this difference for final observables, not only the QED cascade
has to be computed, but first of all, the initial data from Schwinger pair
production has to be determined. For the cascade calculations performed so
far, this is far from being trivial as spatio-temporal dependencies of the
fields have to be accounted for.  The simplest field model often considered
for cascade calculations is a uniformly rotating homogeneous electric
field. Such fields serve as a model for the regions of highest electric field
strength in the antinodes of a circularly polarized standing wave mode, where
QED cascades are expected to occur predominantly. By contrast, first-principles
quantum field theory methods for Schwinger pair production typically allow for
the treatment of constant unidirectional fields
\cite{Heisenberg:1935qt,Schwinger:1951nm} with one-dimensional dependencies on
either space \cite{Kim:2000un,Gies:2005bz,Dunne:2005sx,Dunne:2006st} or time
\cite{Dunne:2006st,Brezin:1970xf,Narozhnyi:1970uv,Popov:1972,Popov:1973az,Gavrilov:1996pz,Dunne:1998ni,Dietrich:2003qf,Piazza:2004sv}.
More involved fields soon lead to an enormous increase in computational
complexity \cite{Dunne:2006ur,Dietrich:2007vw,Ruf:2009zz,Hebenstreit:2011wk}.

In the present work, we consider for the first time Schwinger pair production
in a time-dependent rotating spatially homogeneous electric field, that may
help bringing the quantum field theory studies a substantial step closer to
QED cascade calculations. For this, we use the real-time off-equilibrium DHW
(or simply Wigner) formalism on a mean-field level (neglecting back-reactions
on the electromagnetic field) as developed in
\cite{BialynickiBirula:1991tx}. We demonstrate that the coupled partial
differential equations (PDE) for the components of the Wigner function can be
mapped onto simpler (modified) quantum kinetic equations, as first shown for
unidirectional fields in \cite{Hebenstreit:2010vz,Hebenstreit:2011pm}. This
resemblance to quantum kinetic theory
\cite{Smolyansky:1997fc,Kluger:1998bm,Schmidt:1998vi,Bloch:1999eu,Alkofer:2001ik,Hebenstreit:2008ae}
allows the use of the method of characteristics to solve the PDE with
conventional high-precision algorithms.

This technique gives access to the real-time evolution of the distribution
functions and physical observables such as the asymptotic pair distributions
in momentum space and the total particle yield. We find that the qualitative
features of the latter can be interpreted within semiclassical pictures of
pair production (inspired by atomic ionization)
\cite{Brezin:1970xf,Popov:1972,Dunne:2005sx} using an effective Keldysh
parameter that accounts for the time scales in the problem. By contrast, the
momentum distribution is strongly influenced by the field rotation, giving
rise to structures that can be interpreted in terms of quantum interferences
\cite{Hebenstreit:2009km,Heinzl:2010vg,Dumlu:2010ua,Dumlu:2011rr} and
multiphoton processes. We believe that these characteristic structures can
serve as a seed for a QED cascade, potentially leaving its imprints in the
successive complex many-body dynamics. If so, Schwinger pair production should
be well distinguishable from more conventionally sourced QED cascades in
rotating electric fields.

This work is organized as follows: Section~\ref{sec:wigner} briefly
summarizes the DHW formalism as it is needed for the main part of the
paper. In Sect.~\ref{sec:rotfields} the equations of motion for the
Wigner function are adapted to spatially homogeneous, but
time-dependent electric fields of arbitrary directionality. This leads
to a set of quantum kinetic equations generalizing those previously
studied in the literature to the case of arbitrary time-dependent
field directions. Our quantitative results for a rotating field are
presented in detail in Sect.~\ref{sec:results}. Conclusions are drawn
in Sect.~\ref{sec:concl}. Some useful details of the numerical
implementation are deferred to the Appendix.

%-------------------------------------------------------------------------------
\section{The Wigner Function}\label{sec:wigner}
%-------------------------------------------------------------------------------

The DHW formalism for Schwinger pair production has first been
presented in \cite{BialynickiBirula:1991tx}. Comprehensive summaries
for the present context as well as exact solutions for particular
electric fields can be found in
\cite{Hebenstreit:2010vz,Hebenstreit:2011pm}. In the following, we
give a brief summary of the formalism as needed for the present work,
following the original literature. The starting point is the Wigner
operator defined as the Fourier (Wigner) transform of the equal time
density operator of two Dirac field operators in the Heisenberg
picture,
\begin{widetext}
\begin{equation}
  \label{eqn:WignerDefFT}
  \mathcal{\widehat{W}}_{ab}(t,\vec{x},\vec{p}):=
  -\frac12\int d\vec{s}\,e^{-\frac{i}{\hbar}\vec{p}\cdot\vec{s}}\,
e^{-ie\int_{\vec{x}+\nicefrac{\vec{s}}{2}}^{\vec{x}-\nicefrac{\vec{s}}{2}}\vec{\hat{A}}(t,\vec{x}'\!)\cdot d\vec{x}'}
\commutator{\hat{\Psi}_a(t,\vec{x}+\nicefrac{\vec{s}}{2})}{\hat{\overline{\Psi}}_b(t,\vec{x}-\nicefrac{\vec{s}}{2})},
\end{equation}
\end{widetext}
where the Wilson line operator accounts for gauge invariance. Here
$\vec{x}$ is a center-of-mass coordinate of the underlying two-point
correlator, and $\vec{s}$ denotes a relative coordinate the Fourier
conjugate of which defines the kinetic momentum $\vec{p}$. Replacing
the gauge field operator by an external classical field (corresponding
to a mean-field or Hartree approximation), the vacuum expectation
value of the Wigner operator can be taken which defines the Wigner
function $\mathcal{W}$. Using the equations of motion for the fermionic Heisenberg
operators, the dynamical equation for the Wigner function can be
written as
\begin{equation}
  \label{eqn:Wigner-EoM}
  \Dt \mathcal{W}=-\frac12 \vDx \commutator{\gamma^0\vec{\gamma}}{\mathcal{W}}-im\commutator{\gamma^0}{\mathcal{W}}-i\vec{P}\anticommutator{\gamma^0\vec{\gamma}}{\mathcal{W}}
\end{equation}
with the pseudo-differential operators
\begin{align*}
   \Dt &=\partial_t + e\int_{-\nicefrac12}^{\nicefrac12} d\lambda\,\vec{E}(t,\vec{x}+i\lambda\nablap)\cdot\nablap\,,
\\
  \vDx &=\nablax + e\int_{-\nicefrac12}^{\nicefrac12} d\lambda\,\vec{B}(t,\vec{x}+i\lambda\nablap)\times\nablap\,,
\\
  \vec{P} &=\vec{p}-ie\int_{-\nicefrac12}^{\nicefrac12} d\lambda\,\lambda\vec{B}(t,\vec{x}+i\lambda\nablap)\times\nablap\,,
\end{align*}
Here, we have used the conventions
$\anticommutator{\gamma^\mu}{\gamma^\nu}=+2\eta^{\mu\nu}=+2\operatorname{diag}(1,-1,-1,-1)$
and worked in temporal gauge $A_0=0$.  In the language of Feynman
diagrams, the mean-field approximation corresponds to neglecting
radiative corrections, which is justified by the smallness of the
fine-structure constant $\alpha$.  The Wigner function can be
decomposed in terms of a complete basis of the Clifford algebra,
($\mathbbm{1},\gamma^5,\gamma^\mu,\gamma^\mu\gamma^5,\sigma^{\mu\nu}:=\frac{i}{2}\commutator{\gamma^\mu}{\gamma^\nu}$),
\begin{align}
 \label{eqn:chiral}
 \mathcal{W} = \frac14
      \left(
        \mathbbm{1} \bbs + i \gamma_5 \, \bbp
        +\gamma^\mu \, \bbv_\mu + \gamma^\mu \gamma_5 \, \bba_\mu
        +\sigma^{\mu\nu} \, \bbt_{\mu\nu}
      \right)
\end{align}
with correspondingly transforming coefficient functions, $\bbs,\bbp,\bbv_\mu,\bba_\mu,\bbt_{\mu\nu}$.
The equation of motion can be decomposed accordingly, yielding
\begin{align}
  \begin{rarray}{1.5pt}[c]{rrlll}
      \mathbbm{1}:
      &\qquad\Dt \bbs
      &=
      &+2\vec{P}\cdot\vec{\bbt}^{\,1}
    \\
      i\gamma_5:
      &\Dt \bbp
      &=
      &-2\vec{P}\cdot \vec{\bbt}^{\,2}
      &-{2m \, \bba^0 }
    \\
      \gamma^0:
      &\Dt \bbv^0
      &=-{\vDx\vec{\bbv}}
    \\
      \gamma^0\gamma_5:
      &\Dt \bba^0
      &=-{\vDx\vec{\bba}}
      &
      &+{2m\,\bbp}
    \\
      \gamma^i:
      &\Dt \vec{\bbv}
      &=-{\vDx \bbv^0}
      &-{2\vec{P}\times \vec{\bba} }
      &-{2 m\, \vec{\bbt}^{\,1}}
    \\
      \gamma^i\gamma_5:
      &\Dt \vec{\bba}
      &=-{\vDx \bba^0}
      &-{2\vec{P}\times \vec{\bbv} }
    \\
      \sigma^{0i}:
      &\Dt \vec{\bbt}^{\,1}
      &=-{  \vDx \!\times  \vec{\bbt}^{\,2} }
      &-{2\vec{P}\, \bbs}
      &+{2m\, \vec{\bbv}}
    \\
      \frac12\varepsilon_{ijk}\sigma^{jk}:
      &\Dt \vec{\bbt}^{\,2}
      &=+{ \vDx\! \times \vec{\bbt}^{\,1}}
      &+{2\vec{P}\, \bbp}
  \end{rarray}
\end{align}
with
\begin{align*}
  \left(\vec{\bbt}^{\,1}\right)_i &:= \bbt_{0i}-\bbt_{i0},   &
  \left(\vec{\bbt}^{\,2}\right)_i &:= \epsilon_i^{\phantom{i}jk}\bbt_{jk}\,.
\end{align*}
A solution of these equations requires initial conditions.
For electromagnetic fields that vanish at asymptotically early times, $t\to -\infty$, initial conditions are given by the vacuum solution
\begin{align}
   \quad \bbs_\vac
  &=  \frac{-2m}{\omega(\vec{p}\,)}
  \,,\quad \vec{\bbv}_\vac
  =   \frac{-2\vec{p}}{\omega(\vec{p}\,)}\,,
\end{align}
where $\omega(\vec{p})=\sqrt{m^2+\vec{p}^2}$.  The coefficient
functions can be viewed as distribution functions in phase space with
a pseudo-probabilistic interpretation. A full probabilistic
interpretation in coordinate or momentum space arises upon integrating
over momenta or coordinates, respectively.

An important auxiliary quantity is the combination
\begin{equation}
 \label{eqn:energdens}
 \epsilon(t,\vec{x},\vec{p}) = \vec{p}\cdot \vec{\bbv}(t,\vec{x},\vec{p})+m\,\bbs(t,\vec{x},\vec{p}),
\end{equation}
which can be interpreted as a (phase space) energy density of the fermionic fields.
The corresponding one-particle distribution function in phase space is given by the difference of this energy density as compared to the vacuum energy density, normalized by the particle pair energies,
\begin{equation}
 \label{eqn:oneparticledist}
 f(t,\vec{x},\vec{p}) = \frac{1}{2\omega(\vec{p})}\bigl(
 \epsilon(t,\vec{x},\vec{p}) - \epsilon_\vac(t,\vec{x},\vec{p}) \bigr)\,.
\end{equation}
This distribution function will be of central interest for our
investigations.  It is important to stress that the principles of
quantum field theory guarantee a particle interpretation only at
asymptotically large times. For instance, the total number of
particles ($=$ number of anti-particles $=$ number of pairs) produced
out of the vacuum is given by
\begin{equation}
 \label{eqn:numberofparticles}
n=\lim_{t\to\infty}\int d\Gamma\, f(t,\vec{x}, \vec{p}),
\end{equation}
where $d\Gamma$ denotes the phase space measure.

%-------------------------------------------------------------------------------
\section{Quantum kinetic equations in homogeneous electric fields}\label{sec:rotfields}
%-------------------------------------------------------------------------------

The Wigner formalism, in principle, can deal with pair production in
arbitrary electromagnetic fields. In practice, computational cost
inhibits a straightforward numerical integration of the coupled set of
partial differential equations, see \cite{Hebenstreit:2011wk} for a
first solution in a space- and time-dependent field in 1+1
dimensions. This limitation suggests to concentrate on highly
symmetric or effectively dimensionally reduced configurations. In
fact, in \cite{Hebenstreit:2010vz,Hebenstreit:2011pm} it has been
shown that the Wigner formalism for unidirectional time-dependent
spatially homogeneous electric fields can be mapped onto the simpler
formalism provided by quantum kinetic theory which has mostly been
used in practical computations. In the following, we will drop the
condition of unidirectionality in order to allow for rotating
fields. From a technical viewpoint, this leads to a higher complexity
of the problem in momentum space, but still allows for a mapping of
the Wigner formalism onto a modified quantum kinetic description.

Confining ourselves to spatially homogeneous purely electric fields
($\vec{E}=\vec{E}(t),\,\vec{B}\equiv0$), the equation of motion for
the Wigner function simplifies \cite{BialynickiBirula:1991tx} to
\begin{equation}
\label{eqn:eom-homog}
 \left(
    \partial_t
    +e \vec{E}(t) \cdot \nablap
  \right) \, \bbw
  =  \mathcal{M}\, \bbw,
\end{equation}
with the 10 nontrivial Wigner coefficient functions
\begin{align*}
  \bbw&=\begin{pmatrix}
    \bbs, & \vec{\bbv}, & \vec{\bba}, & \vec{\bbt}^{\,1}\,
  \end{pmatrix}^\intercal, 
   \\
  \text{and }\mathcal{M}
  &=  \begin{pmatrix}
        0 & 0 & 0 & 2\vec{p}^{\,\intercal} \\
        0 & 0 & -2\vec{p}\times &-2m \\
        0 & -2\vec{p}\times & 0 &0 \\
        -2\vec{p} & 2m & 0 & 0
      \end{pmatrix}\,.
\end{align*}
The coefficient functions $\bbp$ and $\vec{\bbt}^{\,2}$ vanish in this
case. Because of spatial homogeneity, all quantities no longer depend
on the space coordinate, but only on time and kinetic momentum, e.\,g.,
$\bbw=\bbw(t,\vec{p})$.

Writing the vacuum solution $\bbw_\vac$
as
\begin{equation}
 \bbw_\vac =\begin{pmatrix} \bbs_\vac, &
\vec{\bbv}_\vac, & \vec{0}, & \vec{0}\, \end{pmatrix}^\intercal=  -2 \bbe_1\,,
\end{equation}
it can straightforwardly be verified that $\bbe_1$ is a unit vector in
$\mathbb{R}^{10}$ with a standard scalar product, $\bbe_1\cdot \bbe_1 =1$.
In the same spirit, the one-particle distribution function as defined
in equation~\eqref{eqn:oneparticledist} can be written as
\begin{equation}
\label{eqn:f-w}
 f = \frac12 \bbe_1 \cdot (\bbw-\bbw_\vac)\,.
\end{equation}
Now, the PDE system~\eqref{eqn:eom-homog} can be converted into an ordinary
differential equation (ODE) system by the method of characteristics,
cf.~\cite{BialynickiBirula:1991tx,Hebenstreit:2010vz,Hebenstreit:2011pm}.
Inserting a specific path $\vec{p}\to\vec{\pi}(t)$ into a PDE of the form
\begin{align}
  \label{eqn:pde}
  \bigl( \partial_t + b_i(t)\partial_{p_i} \bigr)g_j(t,\vec{p}) = h_{jk}(t,\vec{p})g_k(t,\vec{p})
\end{align}
and comparing the result to
\begin{align}
  \label{eqn:chain}
  \frac{d}{dt}g_j(t,\vec{\pi}(t)) = \left[\bigl( \partial_t + (\partial_t\pi_i)\partial_{p_i}\bigr)g_j(t,\vec{p})\right]_{\vec{p}=\vec{\pi}(t)},
\end{align}
we find
\begin{equation}
  b_i(t) = \partial_t\pi_i\,.
\end{equation}
This implies that with
\begin{align}
 \label{eqn:trajectory}
 \vec{\pi}_{\vec{q}}(t) = \int_0^t \vec{b}(t')\,dt' + \vec{\pi}_0+ \vec{q}\,,
\end{align}
and some arbitrary constant $\vec{\pi}_0$ to be fixed below, we can solve
\begin{align}
  \label{eqn:ode}
  \frac{d}{dt}\tilde{g}_j(t,\vec{q}) = \tilde{h}_{jk}(t,\vec{q})\tilde{g}_k(t,\vec{q}),
\end{align}
in order to find the solution to the original equation along the characteristics
$\vec{\pi}_{\vec{q}}(t)$ according to $\tilde{g}_j(t,\vec{q})\equiv
g(t,\vec{\pi}_{\vec{q}}(t))$.  Note that $\vec{q}$ now is only a
parameter and can be freely chosen before solving the equation. For a
complete picture of the solution as a function of $t$ and $\vec{p}$,
we have to solve the equation for several values of $\vec{q}$.

With this method, the equation of motion of the Wigner function can be
brought to the simple form
\begin{equation}
 \label{eqn:eom-ode}
 \dot{\tilde{\bbw}}=\tilde{\mathcal{M}}\tilde{\bbw},
\end{equation}
where the tilde notation indicates as before that the corresponding
functions are evaluated along the path 
$\vec{p}=\vec{\pi}_{\vec{q}}(t)$, and the resulting function is then
understood as a function of $\vec{q}$, e.\,g.,
$\tilde{w}(t,\vec{q})\equiv w(t,\vec{p}=\vec{\pi}_{\vec{q}}(t))$.  For
a suitable choice of $\vec{\pi}_0$, the characteristics
$\vec{\pi}_{\vec{q}}(t)$ coincide with the classical momentum space
paths of electrons in the external field according to
\begin{align}
 \label{eqn:trajectoryspecial}
 \vec{\pi}_{\vec{q}}(t) &= e\int_0^t \vec{E}(t')\,dt' + \vec{\pi}_0+ \vec{q} \nonumber\\
                        &= -e\vec{A}(t)+ \vec{q}\,,
\end{align}
where we have used the Weyl gauge and set $\vec{\pi}_0=-e\vec{A}(0)$
in the second line. With this choice, the
variable $\vec{q}$ can be interpreted as the canonical momentum
whereas $\vec{\pi}_{\vec{q}}(t)$ is the kinetic momentum on a
trajectory.

Also the one-particle distribution function $\tilde{f}$ can be
considered along the characteristics parameterized by $t$ and
$\vec{q}$,
\begin{equation}
\label{eqn:f-traj}
 \tilde{f} = \frac12 \tilde{\bbe}_1 \cdot (\tilde{\bbw}-\tilde{\bbw}_\vac)\,.
\end{equation}
With \Eqref{eqn:eom-ode} our original PDE problem is mapped onto an
ODE problem which is, in principle, amenable to standard solution
methods. Upon insertion of the solution into \Eqref{eqn:f-traj}, we
can extract the physical information. 

However, in practice, the problem is numerically challenging for the
following reason: in a case where only few pairs are created the
solution only differs very slightly from the vacuum solution.  In
order to compute this difference, the numerical solution
of~\Eqref{eqn:eom-ode} must be determined with a very high
precision. This problem is amplified further by the fact that the
projection onto the unit vector $\bbe_1$ also results in the computation
of small differences of numerical quantities.  Thus a reformulation of
the equation of motion is required to obtain better numerical
sensitivity for small pair creation rates.  For simplicity the tilde
symbol is dropped from now on, and all following equations will be
considered along the characteristics.

For our reformulation, we decompose the Wigner functions into a
component parallel to $\bbe_1$ and its orthogonal complement.  The latter
part is parameterized as $\mathcal{T}\bbw_9$ where $\mathcal{T}$ is a
$10\times9$ matrix chosen such that
$\bbe_1^\intercal\mathcal{T}=\mathbbm{o}^\intercal$ (with
$\mathbbm{o}$ being the zero vector). $\mathcal{T}$ must also be
constructed such that a $9\times10$ matrix $\mathcal{R}$ exists,
that satisfies $\mathcal{R}\mathcal{T}=\mathbbm{1}_9$. The quantity
$\bbw_9$ denotes an auxiliary 9 component vector.
We can now plug the ansatz
\begin{align}
 \bbw &= 2(f-1)\, \bbe_1 + \mathcal{T}\bbw_9,
\end{align}
which is clearly compatible with \Eqref{eqn:f-traj}, into the equation of motion.
We obtain
\begin{equation}
 2\dot{f}\bbe_1 + 2(f-1)\,\dot{\bbe}_1 + \dot{\mathcal{{T}}}\bbw_9 + \mathcal{T}\dot{\bbw}_9 =\mathcal{MT}\bbw_9,
\end{equation}
where we have used that $\mathcal{M}\bbe_1\equiv\mathbbm{o}$.  Evolution
equations for $f$ and $\bbw_9$ can finally be read off by
applying $\bbe_1$ and $\mathcal{R}$, respectively, to both sides of
the equation.  The resulting equations read
\begin{align}
\label{eqn:simplf.eom}
 \begin{split}
   \dot{{f}}
  &=  \frac12 \dot{{\bbe}}_1^\intercal\mathcal{{T}} {\bbw}_9\\
  \dot{{\bbw}}_9
  &= {\mathcal{M}}_9{\bbw}_9 + 2(1-{f})\mathcal{{R}}\dot{{\bbe}}_1
 \end{split}
 \intertext{with}
 \mathcal{{M}}_9 &= \mathcal{{R}} \left(\mathcal{{M}} - {\bbe}_1\dot{{\bbe}}_1^\intercal \right) \mathcal{{T}} +\dot{{\mathcal{R}}}\mathcal{{T}},
\end{align}
where $0=\frac{d}{dt}\mathcal{RT}$ has been used.
In our studies we have chosen
\begin{align}
 \mathcal{T} &= \left.\begin{pmatrix}
                   \begin{matrix}
                    -\nicefrac{p_x}{m}&-\nicefrac{p_y}{m}&-\nicefrac{p_z}{m}&{0}&{\cdots}\!&{0}
                   \end{matrix} \\
                   \mathbbm{1}_9
                \end{pmatrix}\right|_{\vec{p}\to\vec{\pi}_{\vec{q}}(t)},
\end{align}
and accordingly
\begin{align}
%   \intertext{and}
  \mathcal{R} &= \begin{pmatrix}
                   \mathbbm{o}
                  &
                  \mathbbm{1}_9
                 \end{pmatrix}\,.
\end{align}
With this choice
\begin{align*}
  \mathcal{M}_9 &=
 \left.
 \begin{pmatrix}
   \mathcal{A} & -2\vec{p}\times & -2m
   \\
   -2\vec{p}\times & 0 & 0
   \\
   \mathcal{B} & 0 & 0
 \end{pmatrix}
 \right\rvert_{\vec{p}=\vec{\pi}_{\vec{q}}(t)},\\
 \quad\text{where}\qquad
 \mathcal{A} &= \frac{-e}{m^2+\vec{p}^{2}} \vec{p} \cdot \vec{E}^\intercal\,,
 \\
 \mathcal{B} &= \frac{2}{m} \left( m^2 \mathbbm{1}_3 + \vec{p}\cdot\vec{p}^{\intercal} \right)\,.
\end{align*}
The matrix $\mathcal{M}_9$ actually has a three dimensional
kernel, indicating a possible redundancy in the Wigner functions for
the present case. However, since the kernel is field- and thus
time-dependent, it is numerically more convenient to solve the
differential equations \eqref{eqn:simplf.eom} without explicitly
projecting out this redundancy.  In this form, the differential
equations -- though still challenging -- are accessible to
straightforward numerical integration. The initial values for the
original equations simply translate into $f=0$ and $\bbw_9=0$ for
$t\to-\infty$. In practice, the numerical integration is not initiated
at negative infinity but at some finite value.  This corresponds to a
discontinuous switch-on of the electric field which leads to
artificial transient oscillations that subside sufficiently fast for
our purposes.  This and some further details of the numerical
evolution of this set of equations are discussed in
App.~\ref{App:numdet}.

%-------------------------------------------------------------------------------
\section{Rotating electric fields}\label{sec:results}
%-------------------------------------------------------------------------------

The formalism as developed in the preceding section works for
spatially homogeneous, but time-dependent electric fields of arbitrary
(time-dependent) directionality. As a specific and important example,
we focus on rotating electric fields from now on. More precisely, we
consider a field pulse of the form
\begin{equation}
  \label{eqn:puls-sauter-rot}
  \vec{E}(t)=\frac{E_0}{\cosh^2\left(\nicefrac{t}{\tau}\right)} \begin{pmatrix}
                    \cos(\Omega t) \\
                    \sin(\Omega t) \\
                    0
                  \end{pmatrix}\,,
\end{equation}
characterized by a maximal field value $E_0$, an angular rotation
frequency $\Omega$ and a pulse duration $\tau$. This field
configuration can be viewed as a model for the field in an anti-node
of a standing wave mode with appropriate circular polarization.

For the discussion, it is useful to introduce the dimensionless parameters
\begin{equation}
\epsilon= \frac{E_0}{E_{\text{cr}}} = \frac{eE_0}{m^2}, \quad \sigma=\Omega\tau,
\end{equation}
where $\epsilon$ measures the maximum field strength in units of the
critical field, and $\sigma$ is a measure for the number of full
rotation cycles within the pulse duration. The dimensionful
parameters will be given in units of the QED scale, i.\,e., the electron
mass scale. For instance, the pulse duration is measured in units of
the Compton time $1/m$ in units where $\hbar=c=1$.

\begin{figure}[t]
 \includegraphics{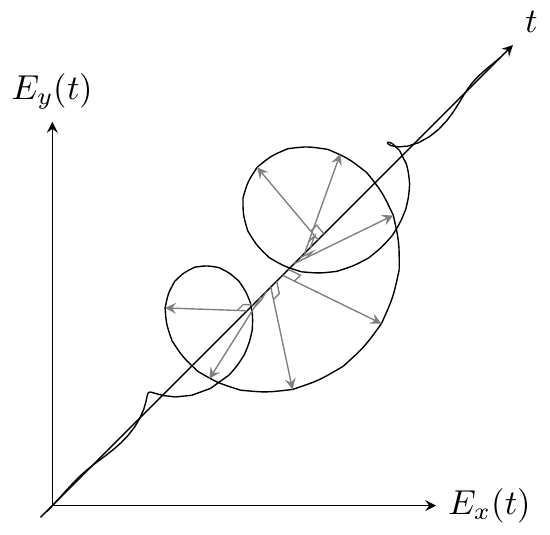}
 \caption{\label{fig:field}Illustration of the rotating electric field
   given in \Eqref{eqn:puls-sauter-rot}.}
\end{figure}

The time evolution of this field is illustrated in
Fig.~\ref{fig:field}. In the limit $\Omega=0$, the rotating field
collapses to a non-rotating Sauter-type field, which is one of the few
examples where the Wigner function can be calculated analytically,
see, e.\,g., \cite{Hebenstreit:2010vz}. Note that a carrier envelope
phase $\phi$ with the replacement $\Omega t \to \Omega t+\phi$ would
have no effect, as it can be transformed to zero by a rotation of the
coordinate system in the $(x,y)$-plane.

%-------------------------------------------------------------------------------
\subsection{Total Particle Yield}
%-------------------------------------------------------------------------------

The total number of pairs per unit volume $\mathcal{N}$ can be
calculated from the distribution function $f$ by integrating over all momenta at
$t\to\infty$, cf. \Eqref{eqn:numberofparticles}, 
\begin{equation}
 \mathcal{N} = \int\!\frac{dq^3}{(2\pi)^3}\,\lim_{t\to\infty}f(t,q)
 = \int\!\frac{dq^3}{(2\pi)^3}\,f_\mathrm{lim}(q)\,.
\end{equation}
In practice, the time integration is stopped at some large but finite
time $t_\text{lim}$. Since the electric field amplitude goes to zero
sufficiently fast for large times, also the time derivative of $f$
vanishes rather rapidly. For the numerical accuracy reached in the
present work, we have observed that values of $t_\text{lim}$ of the
order of $10\tau$ are sufficient, see App.~\ref{App:numdet}.

\begin{figure}[t]
  \includegraphics[width=\columnwidth]{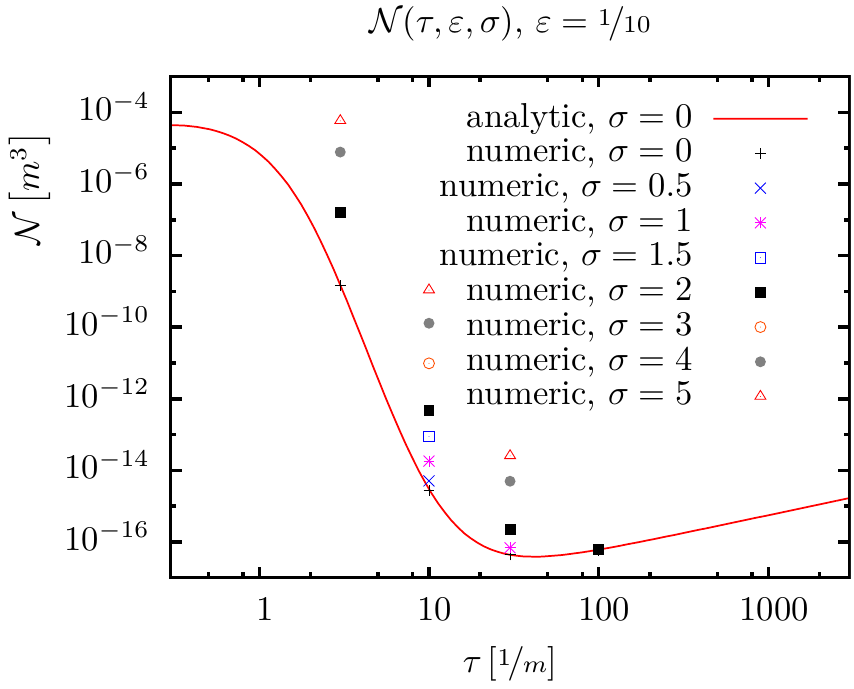}
  \caption{\label{fig:overview}Compilation of the total particle yield
    for a peak field strength $E_0=\varepsilon E_{\text{cr}}$ with
    $\varepsilon=\nicefrac{1}{10}$ for various rotation cycles
    $\sigma$ as a function of the pulse duration $\tau$ in units of
    the Compton time. The solid line marks the analytically soluble
    case of a non-rotating Sauter pulse with $\sigma=0$.}
\end{figure}

As a benchmark, it is useful to compare the particle yield for the
known Sauter type field configuration ($\sigma=0$) with that generated
by rotating fields for different pulse durations and rotation
frequencies. A compilation of results for $E_0=E_{\text{cr}}/10$ and
for various rotation cycles $\sigma = 0, \dots, 5$ as a function of
the pulse duration is depicted in Fig.~\ref{fig:overview}.

Our results from numerically integrating the Wigner
equations~\eqref{eqn:simplf.eom} for various $\sigma$ are displayed as
symbols.  In order to obtain the particle yield $\mathcal{N}$ by
integrating over $\vec{q}$, we have chosen a suitable lattice in
momentum space $\vec{q}$ and calculated $f(t_\mathrm{lim},\vec{q})$
for each lattice point. These results are compared to the analytically
known particle yield for the Sauter pulse for $\sigma=0$
\cite{Hebenstreit:2010vz,Hebenstreit:2011pm} which is shown as
solid/red line. As a check, the numerically obtained results for
$\sigma=0$ (black plus symbols) match the analytic curve adequately
over several orders of magnitude. 

As a main new result, we find that for a given value of pulse duration
$\tau$, the particle yield from a rotating pulse can increase several
orders of magnitude as compared to the non-rotating case.

This phenomenon can be qualitatively understood in simple terms: Let us
first recall that the characteristic shape of the particle yield
$\mathcal{N}(\tau)$ from the Sauter pulse as displayed by the solid curve in
Fig.~\ref{fig:overview} can be discussed in terms of the 
Keldysh adiabaticity parameter,
\begin{equation}
 \label{eqn:gamma1}
 \gamma=\frac{1}{\tau\varepsilon m}= \frac{m}{\tau eE_0}\,.
\end{equation}
For the Sauter pulse, the value of $\gamma$ separates the
nonperturbative Schwinger regime of pair production ($\gamma\ll1$)
from the perturbative multiphoton regime ($\gamma\gg1$). In
particular, the strong increase in the particle yield for
$\tau\lesssim 10$ in Fig.~\ref{fig:overview} (solid curve) is due to
the onset of multiphoton pair production.

Now, switching on the rotation also introduces another time scale $1/\Omega$,
such that the Keldysh parameter defined in \Eqref{eqn:gamma1} is no longer
unique, cf. \cite{Schutzhold:2008pz}. In particular, in the limit of rapid
rotation $\Omega\gg 1/\tau$ we expect the pair production process to be rather
characterized by a Keldysh parameter of the form
\begin{equation}
 \label{eqn:gammamp}
 \gamma^\Omega = \frac{\Omega}{\varepsilon m}\,,
\end{equation}
since the rotation frequency $\Omega$ sets the frequency scale for the
photons collectively dominating the rotating pulse. 

In this sense, the rapid increase of the particle yield for fixed $\tau$ but
increasing $\Omega=\sigma/\tau$ visible in Fig.~\ref{fig:overview} can simply
be interpreted as the onset of multiphoton pair production stimulated by the
photon components of the field with frequency scale $\Omega$ in the
spirit of the folding model of \cite{Nousch:2012xe}. In order to quantify
this simple multiphoton picture, we can define a combined Keldysh parameter
that interpolates between $\gamma$ of \Eqref{eqn:gamma1} and $\gamma^\Omega$
of \Eqref{eqn:gammamp}, for instance,
\begin{equation}
 \label{eqn:gamma2}
 \gamma^* = \frac{\sqrt{1+\sigma^2}}{\tau\varepsilon m}\,.
\end{equation}
We emphasize that this is merely a simple choice and by no means
unique, other interpolating functions with $\gamma^*\to \gamma$ for
$\sigma\to 0$ and $\gamma^* \to \gamma^\Omega$ for $\sigma\to
\infty$ may equally well be used.

\begin{figure}[t]
\includegraphics[width=\columnwidth]{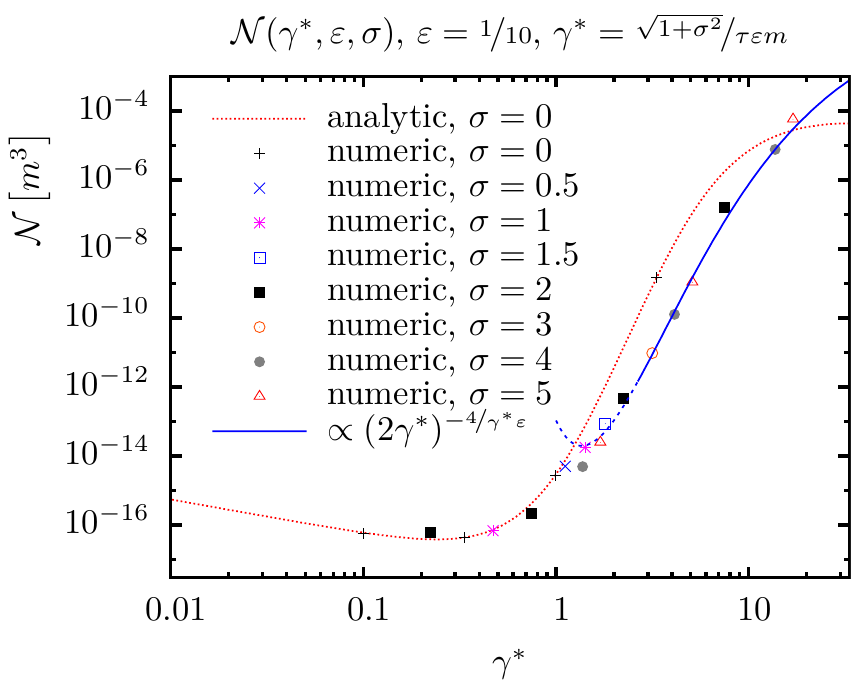}
 \caption{\label{fig:overview2}Particle yield as a function of
   $\gamma^*$ defined in \Eqref{eqn:gamma2} (same data as in
   Fig.~\ref{fig:overview}). The blue solid line shows our simple
   model estimate inspired by a multiphoton picture, matching the data
   on a universal curve for $\sigma\gtrsim1$.}
\end{figure}
Figure~\ref{fig:overview2} now shows the same data as
Fig.~\ref{fig:overview} as a function of $\gamma^*$. We observe that
the data for rotating fields with $\sigma\gtrsim1$ appears to fall on a
universal curve. We conclude that the particle yield no longer depends
on $\sigma$ or $\tau$ individually but rather on a combination which
is approximately given by our choice of $\gamma^*$ in
\Eqref{eqn:gamma2}. The shape of this universal curve can also be
deduced by analogy to the multiphoton regime for linearly polarized
oscillations. According to the semi-classical calculations of Refs.
\cite{Brezin:1970xf,Popov:1972}, the multiphoton pair production rate for fields
of the form $E(t)=E_0\cos(\Omega t)$ with $\gamma^\Omega\gg1$ is given
by
\begin{equation}
 r \propto \left( \frac{eE_0}{2m\Omega}
 \right)^{\nicefrac{4m}{\Omega}} = \left( \frac{1}{2\gamma^\Omega}
 \right)^{\nicefrac{4}{(\varepsilon\gamma^\Omega)}}\,.
\end{equation}
Replacing $\gamma^\Omega$ by $\gamma^*$ and using a fit to this
functional dependence on the Keldysh parameter yields the blue solid
line in Fig.~\ref{fig:overview2} which satisfactorily matches the data
in the multiphoton regime for $\sigma \gtrsim 1$. We conclude that the
universal behavior can indeed be interpreted as a multiphoton effect
induced by the photons at frequency $\Omega$ collectively dominating
the rotating electric field.

%-------------------------------------------------------------------------------
\subsection{Particle Momentum Distribution}
%-------------------------------------------------------------------------------

More information than just the total particle yield is encoded in the
one-particle distribution function $f(t,\vec{p})$. In the present
section, we first study the physically relevant limit of asymptotic
times $t\to t_{\text{lim}}$; more details on the whole time evolution
are discussed in the following subsection.

In order to develop an intuition for the rotating case, let us start
with the simpler case of a Sauter pulse with $\sigma=0$,
cf. \cite{Hebenstreit:2010vz,Hebenstreit:2011pm}. In this case, the
distribution depends only on the kinetic momentum $p_x$ parallel to
the direction of the electric field and the modulus of the perpendicular momentum
$p_\perp=\sqrt{p_y^2+p_z^2}$. The resulting distribution has a single
peak centered at $p_x=\varepsilon\tau$ and $p_\perp=0$.

\begin{figure}[!t]
 \includegraphics{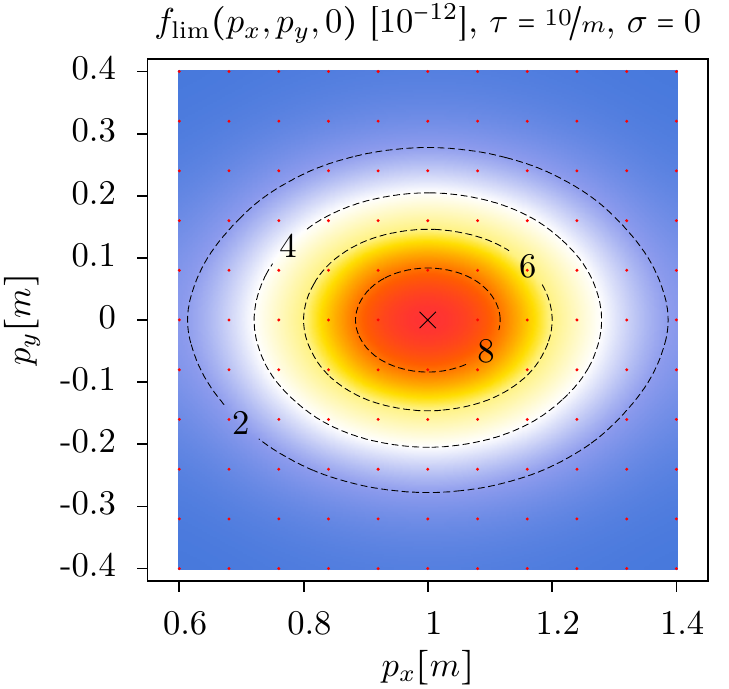}
 \caption{\label{fig:distr.sauter}Particle momentum distribution
   created by a Sauter pulse ($\sigma=0$) with a pulse duration of
   $\tau=\nicefrac{10}{m}$ and a peak field strength
   $\varepsilon=E_0/E_{\text{cr}}=0.1$.}
\end{figure}

An Example for a field strength of $\varepsilon=0.1$ is shown in
Fig.~\ref{fig:distr.sauter} in terms of a contour plot in the
$p_x,p_y$ plane (which is identical with the $p_x,p_\perp$ plane for
$p_z=0$). The quantitative values of the distribution function are
given as the numbers on the contour lines expressed in units given in
brackets in the plot title ($10^{-12}$ in this case). The axes ranges
are chosen such that the plot is centered about the main features of
the distribution while maintaining a square aspect ratio, i.\,e.,
horizontal and vertical axis have the same absolute scale. The contour
plot is produced from a smooth interpolating function; the red dots
indicate the grid of the numerically calculated data points.

The black cross marks the point where the canonical momentum vanishes
$\vec{q}=0$. For the Sauter pulse, this is equivalent to
$\vec{p}=eE_0\tau\,\vec{e}_x$, cf. \Eqref{eqn:trajectoryspecial},
which is the position of the maximum of the
peak.  This is in agreement with a simple semi-classical picture that
particles are predominantly created at rest through instantaneous tunneling
around the time when the field has reached its maximum and are afterwards
accelerated by the electric field to finite momentum.

\begin{figure}[t]
 \includegraphics{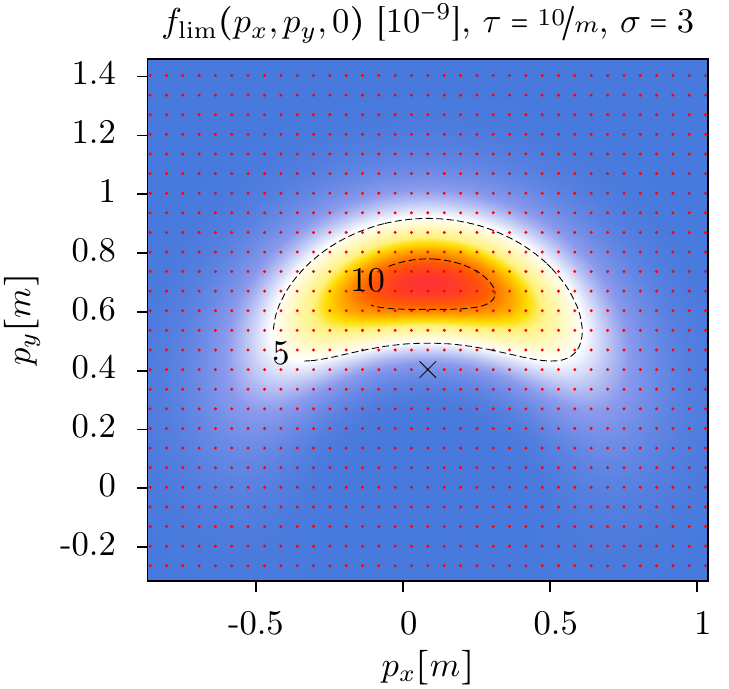}
 \caption{\label{fig:distr.10-3}Momentum distribution of pairs created
   by a rotating electric field pulse with $\sigma=3$ rotation cycles,
   a pulse duration of $\tau=\nicefrac{10}{m}$ and a peak field
   strength $\varepsilon=E_0/E_{\text{cr}}=0.1$.}
\end{figure}

Now, let us switch on rotation: for a moderately small number of
rotation cycles, $\sigma\sim\mathcal{O}(1)$, we observe that the
distribution in momentum space is circularly distorted, see
Fig.~\ref{fig:distr.10-3} for $\sigma=3$. 
In the semi-classical
picture, this can be understood from the fact that the field changes
direction during the creation of particles. As a consequence, the
particles are accelerated into different directions depending on their
instant of creation. In agreement with this simple picture, we also
find that the amount of circular distortion depends on the pulse
length. For instance, decreasing the pulse duration for fixed rotation
frequencies also the circular distortion is weakened, as the created
particles have little time to follow the rotating field
pulse.  We also observe that the peak
of the distribution is no longer given by the point of vanishing
canonical momentum $\vec{q}=0$ (black cross in
Fig.~\ref{fig:distr.10-3}), but at least remains in the vicinity of
this point.
 
It should be stressed that the semi-classical picture cannot cover
all aspects of pair production. Though it is useful for understanding
the asymptotic momentum distributions for the field configurations
studied so far, it finds its limitations in the next example. Even
worse, it seems rather useless for understanding the evolution of the
distribution function at finite times as discussed below in
Subsect.~\ref{sec:time}.

\begin{figure}[!t]
 \includegraphics{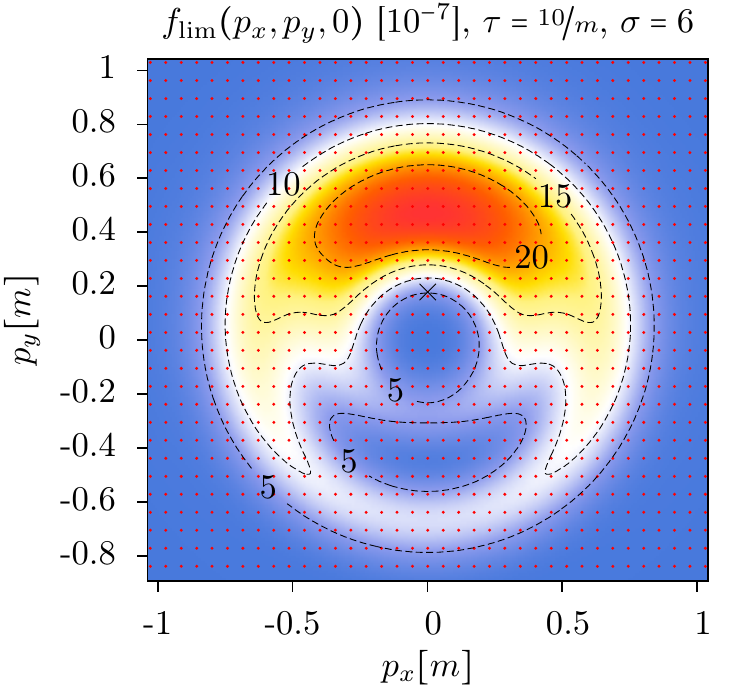}
 \caption{\label{fig:distr.10-6}Momentum distribution of pairs created
   by a rotating electric field pulse with $\sigma=6$ rotation cycles,
   a pulse duration of $\tau=\nicefrac{10}{m}$ and a peak field
   strength $\varepsilon=E_0/E_{\text{cr}}=0.1$.}
\end{figure}

Let us now study larger rotation frequency, i.\,e. a larger number of
rotation cycles $\sigma$, while keeping the pulse duration fixed. The
case of $\sigma=6$ is shown in Fig.~\ref{fig:distr.10-6}.  For
increasing $\sigma$, we observe that the circular distortion closes
into a ring, and another ring can form inside the first ring, while
the momentum space radius of the outer ring grows. 

An understanding of this pattern requires to go beyond the
quasi-classical particle picture. In fact, the production process is a
quantum mechanical process, implying that the produced particles also
carry phase information. In particular, the right-moving particles
with positive $p_x$ carry a generically different phase than the
left-moving particles with negative $p_x$ for the case of a
circularly distorted distribution. For larger $\sigma$, these ends of
the distribution now meet again in momentum space to form a
ring. Since this corresponds to a sum of quantum mechanical wave
functions with different phases, we expect to see an interference
pattern. This is in fact, what we observe along the negative $p_y$
axis at $p_x=0$.

We can even go one step further and try to relate the maxima of the
ring-shaped interference pattern to the excess energy in a multiphoton
pair production process: a multiphoton picture for pair production
implies that a merger of $n$ photons of frequency $\Omega$ distributed
over the electron and positron should give an average excess energy of
$p=\frac{1}{2}\sqrt{(n\Omega)^2-4m^2}$ per particle.  In fact, for
example the outer ring in Fig.~\ref{fig:distr.10-6} has a radius of
about $p\simeq0.66m$ which approximately matches with the current
pulse parameters for $n=4$,
$p=\sqrt{\left({2\cdot0.6m}\right)^2-m^2}$. Slight deviations to this
  simple picture can be interpreted as signatures of the effective mass of the
  fermion in the strong-field environment \cite{Kohlfurst:2013ura}.

\begin{figure}[!t] 
 \includegraphics{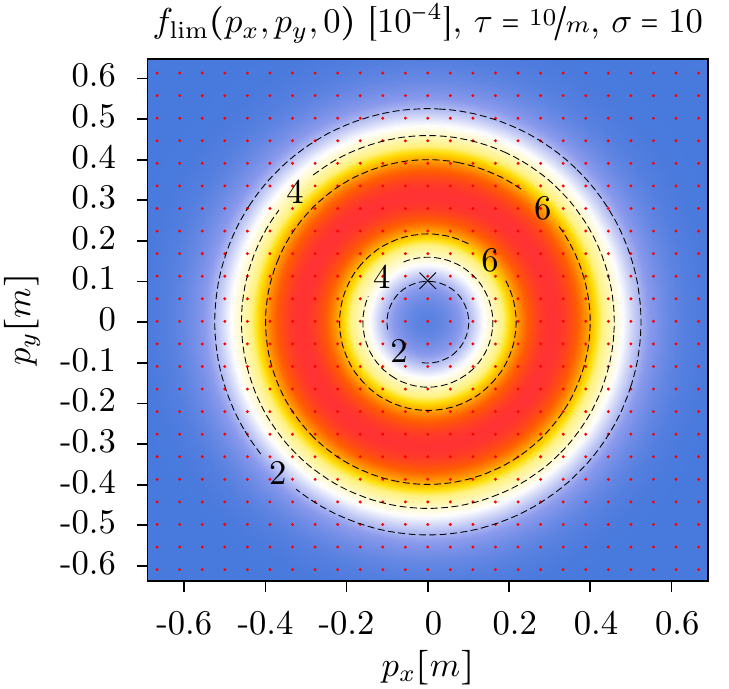}
 \caption{\label{fig:distr.10-10}Momentum distribution of pairs created
   by a rotating electric field pulse with $\sigma=10$ rotation cycles,
   a pulse duration of $\tau=\nicefrac{10}{m}$ and a peak field
   strength $\varepsilon=E_0/E_{\text{cr}}=0.1$.}
\end{figure}

As can be anticipated, the particle distribution becomes isotropic for
very high rotation frequencies, because the field rotates many times
under the envelope and does not change much from one cycle to another.
We observe this behavior already for $\sigma=10$ for our pulse
parameters, as is depicted in Fig.~\ref{fig:distr.10-10}.

We also observe that all distributions exhibit a mirror symmetry in
momentum space with respect to $q_x=0$.  This is already the case for
the Sauter pulse and holds still true for the more involved momentum
distributions generated from the rotating electric field. At first
sight, this looks surprising, because the electric field at the
instant of reaching its peak field strength points into the $x$
direction, seemingly representing an asymmetry in this direction.  As
shown in \cite{Dumlu:2011rr} for the case of the Sauter pulse, the
observed symmetry can be related to the time reversal symmetry of the
$x$ component of the electric field. The very same arguments go
through for our choice of the rotating electric field and hence imply
the same symmetry property for the momentum space distribution. 

From a phenomenological viewpoint, this leads us to an important
conclusion: for rotating electric field pulses with rotation cycles on
the order of $\sigma \gtrsim \mathcal{O}(1)$, we observe a
characteristic particle distribution in momentum space starting from
moderately circularly distorted distributions as in
Fig.~\ref{fig:distr.10-3} to highly asymmetric circular distributions
as in Fig.~\ref{fig:distr.10-6}. Most importantly, the dominant peak
of the distribution does not point into the direction of the peak
electric field ($x$ direction in the present example), but
orthogonally to this direction within the plane of rotation. We
believe that this circular asymmetry can serve as a characteristic
signature for pair production.

We should mention that for increasing pulse durations two numerical problems
arise.  The first problem is an increase of the required computation time per
trajectory.  This is due to oscillations of some components of the function
$\bbw_9$ which demand for the number of
integration steps being at least proportional to the pulse duration.  As a
second problem, the momentum distribution develops filigree structures that
require a large number of lattice points. Both problems increase the necessary
computation time but can still be compensated by using faster computers.

%-------------------------------------------------------------------------------
\subsection{Time Evolution}\label{sec:time}
%-------------------------------------------------------------------------------

\begin{figure}[!t]
 \includegraphics{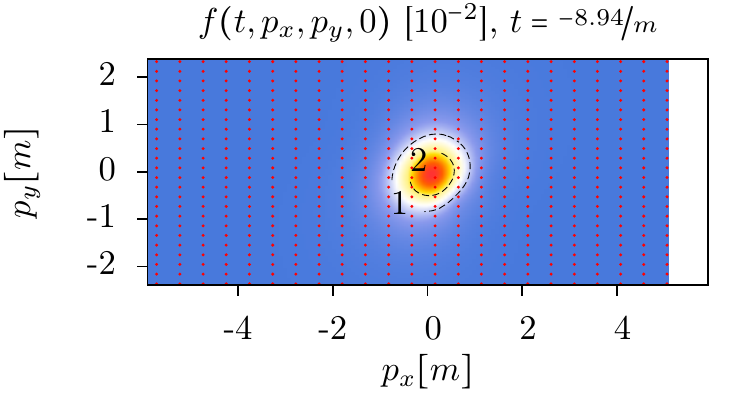}\\%
 \vspace{0.25cm}%
 \includegraphics{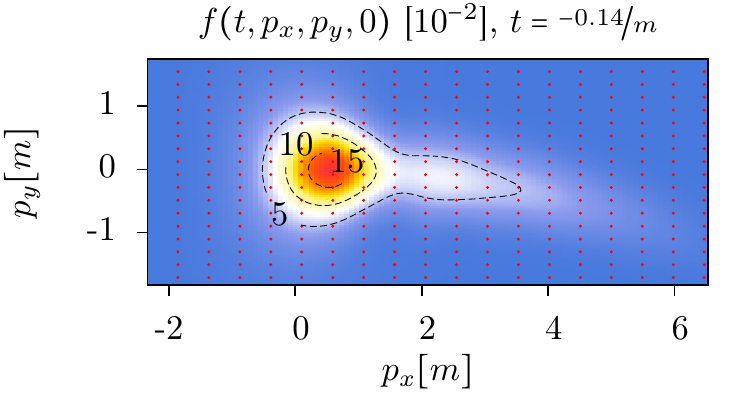}\\%
 \vspace{0.25cm}%
 \includegraphics{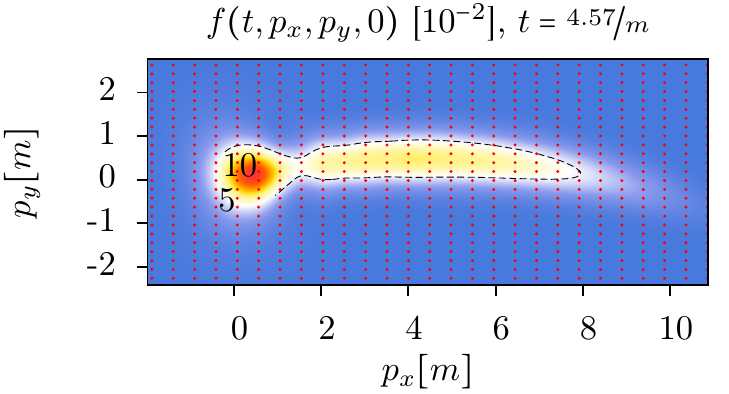}\\%
 \vspace{0.25cm}%
 \includegraphics{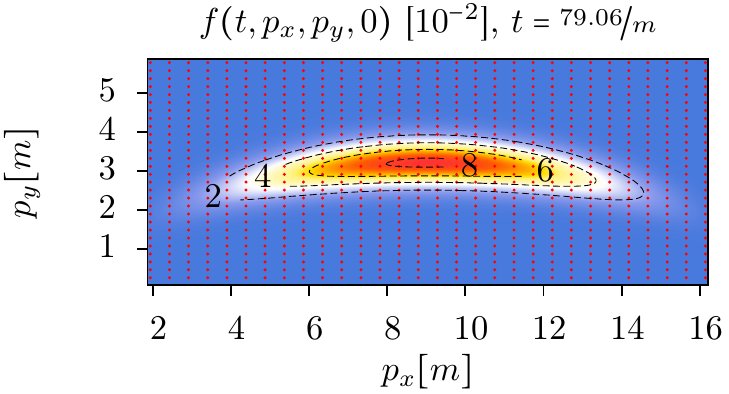}%
 \caption{\label{fig:zeitentw-eps1-sig0.5}Time evolution of the
   distribution function for a strong field pulse with $\varepsilon=1$
   and pulse duration $\tau=\nicefrac{10}{m}$ at $\sigma=\nicefrac12$
   rotation cycles.}
\end{figure}

As emphasized above, the one-particle distribution function can be
interpreted as a phase space particle density only at asymptotic times
$t\to\infty$. At intermediate times, it has no physical meaning in
terms of directly accessible observables. Still, from
\Eqref{eqn:oneparticledist}, the distribution function may give some
intuition on a kind of normalized energy density that may be
attributed to virtual excitations of the electron-positron field.  This
viewpoint is also suggested within quantum kinetic theory where the
Vlasov equation for the distribution function is derived with the aid
of a Bogoliubov transformation. The latter connects vacuum Fock space
ladder operators with quasi-particle creation and annihilation
operators at intermediate times. Only for asymptotic times when the
field has subsided, these quasi-particle operators can be uniquely
connected with a particle Fock space.
Nonetheless the time evolution of the function $f$ continuously
connects the asymptotic regimes and hence carries relevant information
that is worthwhile to be studied. It should, however, be kept in mind
that virtual and real particle information is intertwined at
intermediate times. In the strong field regime, this interlocking
seems less severe such that the time evolution of the distribution
function can be interpreted straightforwardly. As an example, we
consider a pulse with field strength $E_0=E_{\text{cr}}$, i.\,e.,
$\varepsilon=1$, pulse duration $\tau=10/m$ and $\sigma=1/2$ rotation
cycles. A series of snapshots at intermediate times shortly before,
during and after the pulse are shown in
Fig.~\ref{fig:zeitentw-eps1-sig0.5}.
When the pulse starts to set in
(top panel, $t=-8.94/m$), virtual field excitations can be observed
near zero momentum $\vec{p}=0$. Whereas this zero-momentum peak is
further enhanced during the pulse evolution, some part of these
excitations is accelerated along the direction of the field
(predominantly $\sim x$ direction). The distribution develops a tail
along the corresponding momentum direction that acquires a slight
circular distortion due to the field rotation, cf. 2nd and 3rd panel
at $t=-0.14/m$ and $t=4.57/m$ respectively. Towards asymptotic times,
the previously dominant zero momentum peak of virtual excitations
vanishes again, and the momentum-space distribution of outgoing real
particles forms with the characteristic circular distortion visible in
the lowest panel at $t=79.06/m$. (Apart from this distortion, the
result is very similar to that for the Sauter pulse for $\sigma=0$.)

It is interesting to note that the aforementioned mirror symmetry of
the asymptotic distribution about the $q_x=0$ axis is not present at
intermediate times.  This can be explained with the fact that the
$x$ component of the field is not time reversal symmetric if the
evolution is cut off at some finite time $t$. Only the global field
features the required properties.

\begin{figure*}[!t]
\includegraphics{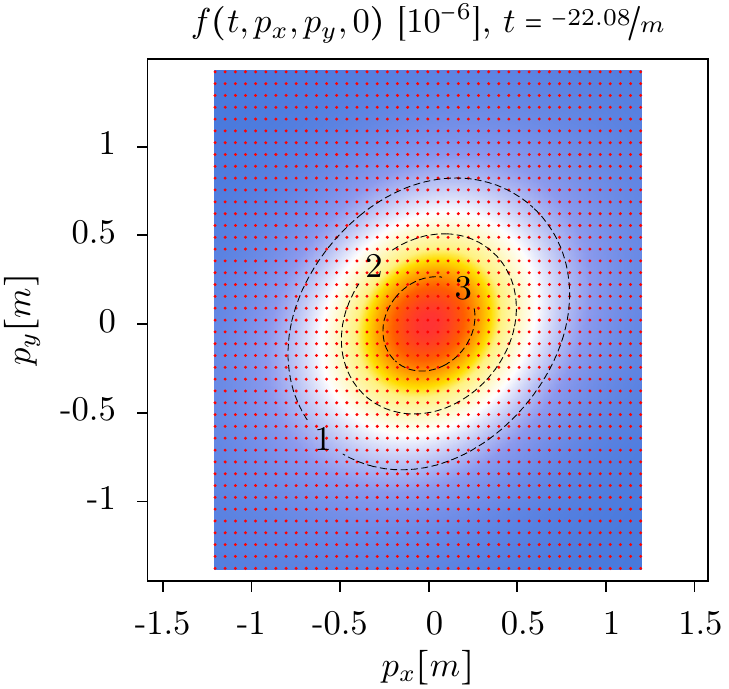}%
 \hspace{0.5cm}%
 \includegraphics{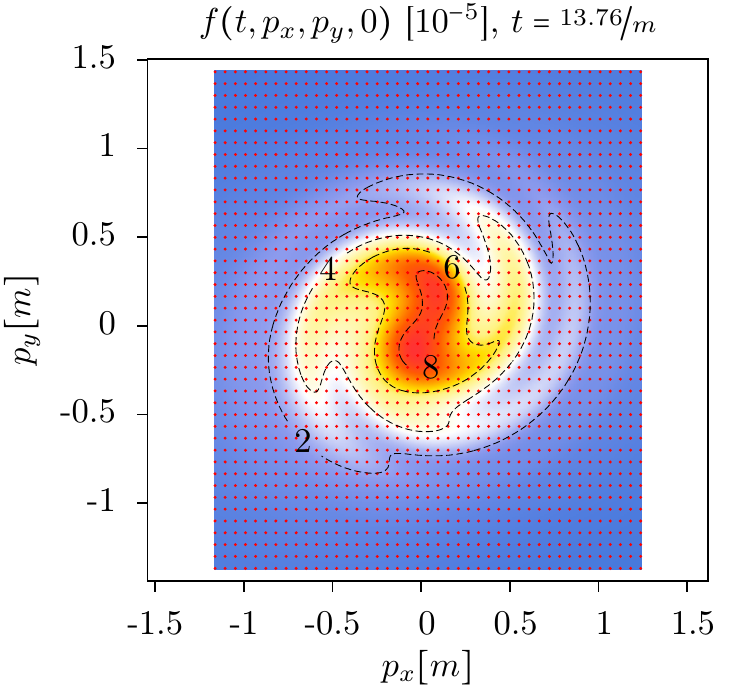}\\
 \vspace{0.5cm}
 \includegraphics{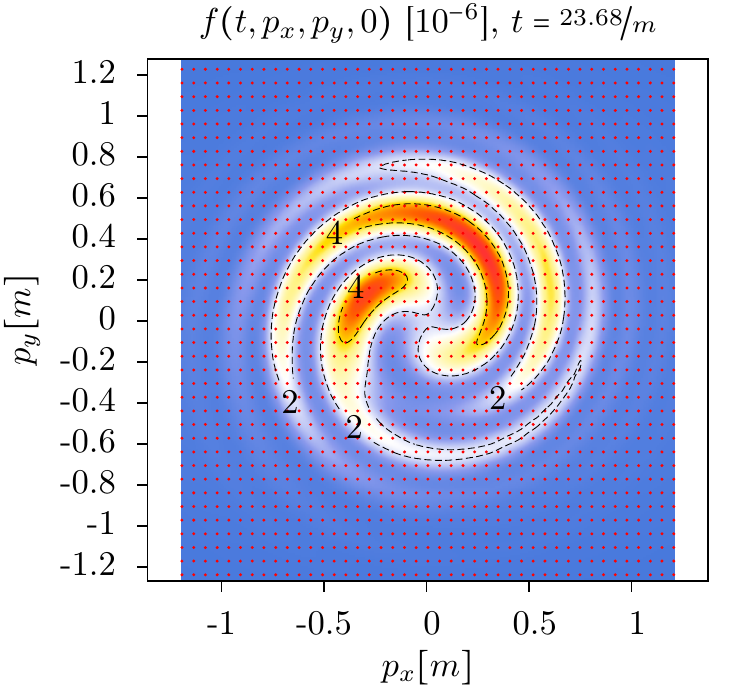}%
 \hspace{0.5cm}%
 \includegraphics{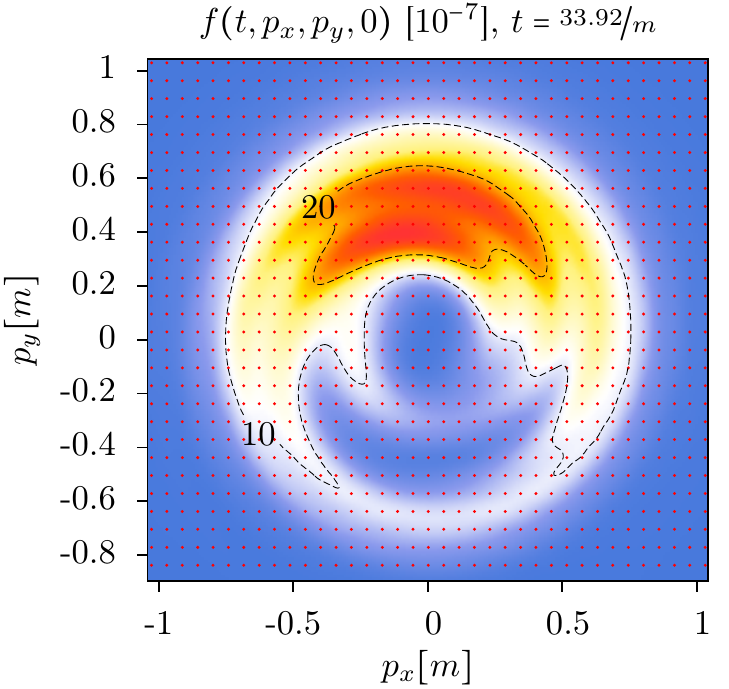}
\caption{\label{fig:zeitentw-f-10-6} Time evolution of the distribution
  function leading to the momentum distribution shown in
  Fig.~\ref{fig:distr.10-6} resulting from a rotating pulse with
  $\varepsilon=\nicefrac{1}{10}$, $\tau=\nicefrac{10}{m}$ and
  $\sigma=6$.}  
\end{figure*}

The separation between virtual excitations and real pairs becomes less
obvious at weaker field strengths. Quantum mechanical phase and
interference effects become more important in this regime. This is
illustrated in Fig.~\ref{fig:zeitentw-f-10-6} for a field strength of
$\varepsilon=E_0/E_{\text{cr}}=0.1$ and a pulse duration of
$\tau=10/m$ with $\sigma=6$ rotation cycles.
There is no clear separation between a virtual excitation peak or a
real particle tail neither in height nor in momentum space. In
addition, the rotation of the field appears to induce stronger
interference effects. Only when the pulse has almost faded away the
virtual excitations relax and the circular distribution pattern starts
to form yielding the asymptotic result of Fig.~\ref{fig:distr.10-6}.

%-------------------------------------------------------------------------------
\section{Conclusion and Outlook}\label{sec:concl}
%-------------------------------------------------------------------------------

We have investigated Schwinger pair production in rotating
time-dependent electric fields for the first time using the DHW
formalism for a numerical computation of the Wigner function. With
this method, we have access to particle/antiparticle distributions in
momentum space, the total particle yield as well as the whole
time-evolution of the production process. 

We find that rotation generically enhances pair production in
comparison with a linearly polarized field pulse. This is
heuristically clear since rotation introduces more Fourier modes that
lead to a pair production increase similar to multiphoton effects for
linear oscillating pulses. We have corroborated this interpretation by
introducing an effective Keldysh parameter that accounts for the time
scales of both the pulse as well as of the rotation. For a
sufficiently large number of rotation cycles, the data for the total
particle yield falls on a universal curve. This universal curve can be
parameterized by a multiphoton description in terms of the effective
Keldysh parameter.

Even more interesting features can be read off from the momentum space
distribution of the produced particles. The rotating field leaves several
characteristic imprints on the distribution depending on the number of
rotation cycles. Rotation can in particular be read off from the shape of the
distribution, from the (partly counterintuitive) location of the peak, and
from resulting interference patterns. Depending on the parameter regime, some
of these features can be understood in a simple quasi-classical picture or in
terms of multiphoton physics, whereas an understanding of the interferences
requires a full quantitative treatment of the quantum mechanical phase
information.

From a technical viewpoint, we have shown that the DHW equations can
be solved with the method of characteristics for general spatially
homogeneous time-dependent electric fields. In particular, a commonly
made restriction to unidirectional fields is not necessary. For this
generalized class of fields, the PDE system of DHW equations can be
mapped onto an ODE system similar to quantum kinetic theory.

In view of our original motivation arising from QED cascades, we
believe that the characteristic momentum space patterns of the
distribution function can serve as a decisive fingerprint of Schwinger
pair production, provided the QED successive cascade preserves a
remnant of this pattern in the final state distributions of either
electrons, positrons or photons. If so, a QED cascade seeded by the
pair production process studied in this work could be quantitatively
distinguished from a more mundanely sourced cascade stimulated by
isotropic vacuum impurities. To answer this question, our results for
the particle distribution should be used as initial conditions of a
QED cascade calculation.

\acknowledgments

We are grateful to R. Alkofer, J. Borchardt, C. Kohlfürst, S. Krause,
H. Ruhl, D. Schinkel and N. Seegert for interesting and enlightening
discussions. We acknowledge support by the DFG under grants Gi 328/5-2
(Heisenberg program), GRK 1523/2, and SFB-TR18. \

\appendix

%-------------------------------------------------------------------------------
\section{Numerical Details}\label{App:numdet}
%-------------------------------------------------------------------------------
% \begin{figure}[t]
%  \includegraphics[width=\columnwidth]{trajektorie}
%  \caption{\label{fig:traj}Distribution function $f$ along an example trajectory.}
% \end{figure}
% 
% \begin{figure}[t]
%  \def\svgwidth{\columnwidth}
%  {\footnotesize
%  \input{pictures//confirm-analytic-plot1-clip.eps_tex}
%  }
%  \caption{\label{fig:comp}Comparison between the analytic solution for
%   the Sauter pulse ($\Omega=0$) (full surface) and the numerical solution along some
%    (colored) trajectories. }
% \end{figure}

\begin{figure*}[t]%
\begin{minipage}[t]{\columnwidth}%
\includegraphics[width=\columnwidth]{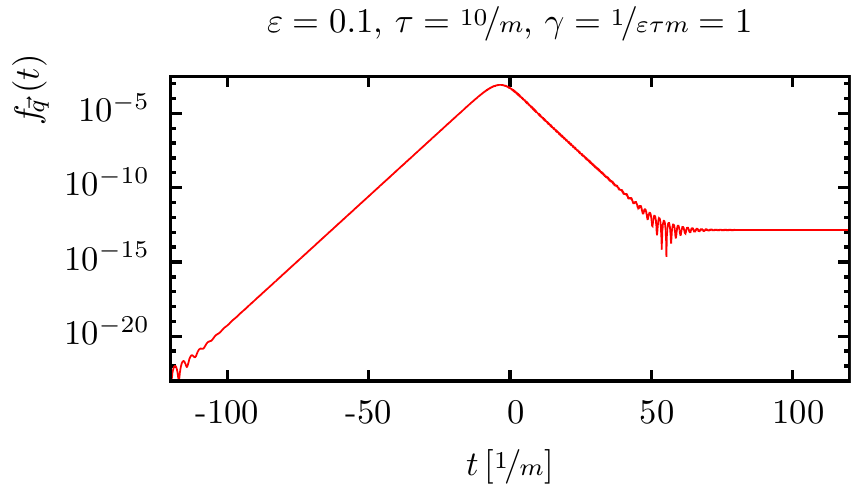}%
 \captionof{figure}{\label{fig:traj}Distribution function $f$ along an example trajectory.}
\end{minipage}%
\hspace{\columnsep}%
\begin{minipage}[t]{\columnwidth}%
 \def\svgwidth{\columnwidth}%
 {\footnotesize%
 \ifpdf%
  \input{pictures//confirm-analytic-plot1-clip.pdf_tex}%
 \else%
  \input{pictures//confirm-analytic-plot1-clip.eps_tex}%
 \fi%
 }%
 \captionof{figure}{\label{fig:comp}Comparison between the analytic solution for
  the Sauter pulse ($\Omega=0$) (full surface) and the numerical solution along some
   (colored) trajectories.}
\end{minipage}
\end{figure*}

Numerical solutions of equation~\eqref{eqn:simplf.eom} have been
computed using \textit{Wolfram Mathematica}.  It's built-in numerical
differential equation solver \texttt{NDSolve} uses various solution
methods in combination with adaptive step size and error estimation
techniques.  The main method used for ordinary differential equations
is \texttt{LSODA} which is part of the \texttt{ODEPACK} package
\cite{Hindmarsh:1983}.  It dynamically switches between nonstiff
(Adams) and stiff (Backward differentiation formula) methods.

In order to achieve the required accuracy it is important to set the
error bounds for the adaptive step size.  The \texttt{NDSolve} method
of \textit{Mathematica} accepts two parameters \texttt{AccuracyGoal} and
\texttt{PrecisionGoal}.  If a function $f$ is calculated from some
ODE the resulting upper bound for the estimate integration
error is
\begin{equation}
 \Delta_\mathrm{max} = 10^{-\mathtt{AccuracyGoal}} + \lvert f\rvert\, 10^{-\mathtt{PrecisionGoal}}\,,
\end{equation}
where $10^{-\mathtt{AccuracyGoal}}$ is a tolerable absolute error and
$10^{-\mathtt{PrecisionGoal}}$ is a tolerable relative error.
The value of the actual function $f$ ranges over many orders of
magnitude as shown in Fig.~\ref{fig:traj} for an example trajectory
and a typical set of parameters. The result of physical interest is
the comparatively small function value at asymptotically large times.
As the peak value near $t=0$ exceeds the asymptotic result by many
orders of magnitude, relative errors are generically unacceptable,
because the error bound could then easily exceed the final physical
value during time steps near the peak value. We hence accept only an
absolute error by setting \texttt{PrecisionGoal} to the special value
\texttt{Infinity}.  

Also the decay of the transient oscillations caused by actually
switching on the field at a finite (large negative) time can be seen
in Fig.~\ref{fig:traj}.

As shown in Fig.~\ref{fig:comp}, the employed methods can reproduce
analytical solutions quite well. In this plot, the exact solution for
the Sauter pulse with $\Omega=0$ and $\tau=\nicefrac{10}{m}$ and $\varepsilon=1$
shown as a surface plot is superposed by the
numerical solutions along a set of characteristics (colored line
plots). Both solutions agree very well even in the strongly
oscillatory regime near the time of peak field strength $t=0$.%

%-------------------------------------------------------------------------------

%-------------------------------------------------------------------------------

%-------------------------------------------------------------------------------
\end{document}